\documentclass[twocolumn,superscriptaddress,preprintnumbers,amsmath,amssymb,pra]{revtex4}
\usepackage{graphicx}
\UseRawInputEncoding

\begin{document}

\thispagestyle{empty}

\title{Casimir-Polder attraction and repulsion between nanoparticles and
graphene in out-of-thermal-equilibrium conditions
}

\author{
G.~L.~Klimchitskaya}
\affiliation{Central Astronomical Observatory at Pulkovo of the
Russian Academy of Sciences, Saint Petersburg,
196140, Russia}
\affiliation{Peter the Great Saint Petersburg
Polytechnic University, Saint Petersburg, 195251, Russia}

\author{
V.~M.~Mostepanenko}
\affiliation{Central Astronomical Observatory at Pulkovo of the
Russian Academy of Sciences, Saint Petersburg,
196140, Russia}
\affiliation{Peter the Great Saint Petersburg
Polytechnic University, Saint Petersburg, 195251, Russia}
\affiliation{Kazan Federal University, Kazan, 420008, Russia}

\author{
O.~Yu.~Tsybin}
\affiliation{Peter the Great Saint Petersburg
Polytechnic University, Saint Petersburg, 195251, Russia}

\begin{abstract}
The nonequilibrium Casimir-Polder force between a
nanoparticle and a graphene sheet kept at different temperatures is
investigated in the framework of Dirac model using the formalism of
the polarization tensor. It is shown that the force magnitude
increases with increasing temperature of a graphene sheet. At larger
separations an
impact of nonequilibrium conditions on the force becomes smaller.
According to our results, the attractive Casimir-Polder force
vanishes at some definite nanoparticle-graphene separation and
becomes repulsive at larger separations if the temperature of a
graphene sheet is smaller than that of the environment. This effect
may find applications both in fundamental investigations of graphene
and for the control of forces in microdevices of bioelectronics.
\end{abstract}

\maketitle

\newcommand{\ve}{{\varepsilon}}
\newcommand{\ef}{{F_{\rm eq}}}
\newcommand{\tef}{{\tilde{F}_{\rm eq}}}
\newcommand{\nf}{{F_{\rm neq}}}
\newcommand{\at}{{(a,T_E,T_g)}}
\newcommand{\kb}{{k_{\bot}}}
\newcommand{\skb}{{k_{\bot}^2}}
\newcommand{\ot}{{(\omega,k_{\bot};T)}}
\newcommand{\pot}{{(\omega,k_{\bot},T)}}
\newcommand{\ok}{{(\omega,k_{\bot})}}

\section{Introduction}

The Casimir-Polder force \cite{1} acts between two
electrically neutral polarizable particles spaced at a distance far
exceeding their sizes or experienced by the neutral polarizable
particle which is well off a macroscopic interface.
This is an attractive force
caused by the combined action of the zero-point and thermal fluctuations
of the electromagnetic field. In the condition of thermal equilibrium,
i.e., under equal temperatures of the particles, material surface, and
the environment, the Casimir-Polder free energy and force are expressed
via the dynamic polarizability of these particles (atoms) and the
reflection coefficients of electromagnetic fluctuations on the surface
in the framework of the Lifshitz theory \cite{2,3}.
The respective expressions follow from the Lifshitz formula for the Casimir
force between two parallel plates when one of them is treated as a rarefied medium.

The obtained results found numerous applications in both fundamental
and applied physics (see Refs. \cite{4,5} for a review). They were
generalized \cite{6,7,8,9,10,11} for out-of-thermal-equilibrium conditions,
e.g., for the case when the surface is kept at one temperature whereas
a nanoparticle or an atom and the environment are characterized by some
other temperature. Recently the Casimir force out of thermal equilibrium
was considered for two similar plates with temperature-dependent
dielectric permittivity \cite{12} and for two superconducting plates
\cite{13}. The possibility of nonequilibrium repulsive Casimir
force between two parallel plates was demonstrated in Ref. \cite{14}.

The Casimir-Polder force between a nanoparticle and a flat surface is an
important contribution to the total particle-surface interaction which
includes also Born repulsion and mechanical contact forces \cite{15,16}.
Investigation of interaction between nanoparticles and material surfaces
is of great concern for designing sensing devices, such as electrochemical
sensors and biosensors for the needs of bioelectronics \cite{17,18,18a,18b}.
The out-of-thermal-equilibrium Casimir-Polder forces between a small
sphere and a plate and between two small spheres were studies in
Ref.~\cite{20a}  in the framework of general scattering formalism.
It was shown that the force can be both attractive and repulsive and it turns
into zero at some separations.
It was also shown \cite{19} that in out-of-thermal-equilibrium conditions a
nanoparticle in the vicinity of a substrate made of nonreciprocal
plasmonic material may even experience the lateral Casimir-Polder force.

During the last few years considerable study has been given to graphene
which is an one-atom-thick sheet of carbon atoms possessing unusual
properties \cite{20,21,22}. The most important of them is that at
energies below approximately 3 eV \cite{23} the electronic
quasiparticles in graphene are massless and are described by the
relativistic Dirac equation, rather than by Schr\"{o}dinger equation,
where the speed of light $c$ is replaced with the Fermi velocity
$v_F\approx c$/300. The Casimir-Polder interaction of different atoms
with graphene in thermal equilibrium was studied in
Refs. \cite{24,25,26,27,29,30,31,32,33,34,37,39}.
Interaction of nanoparticles with graphene also attracted much recent
attention \cite{40,41,42,43,44}, in particular, for graphene on lipid
membranes, taking into consideration prospective applications in
bioelectronics \cite{45,46,47,48,48a}.

In this paper, we investigate the Casimir-Polder force between
nanoparticles and the freestanding graphene sheet in
out-of-thermal-equilibrium conditions. In doing so, the temperature
of graphene $T_g$ may be either higher or lower than the temperature
of nanoparticles which is assumed to be equal to the environmental
temperature $T_E$. The electromagnetic response of graphene is
described on the basis of first principles of quantum electrodynamics
at nonzero temperature using the polarization tensor in
(2+1)-dimensional space-time \cite{49,50,51,52}. An important point is
that the response functions of graphene strongly depend on temperature.
This leads to an unexpectedly large thermal effect in the Casimir
force from graphene at short separations even in the state of thermal
equilibrium \cite{53,54,55}. We show that with increasing $T_g$ the
magnitude of graphene-nanoparticle force increases. For $T_g > T_E$
the nonequilibrium Casimir-Polder force remains attractive. However,
according to our results, for $T_g < T_E$ the force between a
nanoparticle and a graphene sheet vanishes at some separation distance
and becomes repulsive at larger separations.
Unlike Ref.~\cite{20a}, where the response functions of interacting
bodies are temperature-independent and nonequilibrium effects at short
separations are negligible, in our case they become large for nanoparticles
remote from graphene for only a few hundred nanometers.
Possible applications of these results are discussed.

\section{Nonequilibrium Casimir-Polder force between a nanoparticle and
a graphene sheet}

We consider the dielectric or metallic spherical
nanoparticles of radius $R$ at the environmental temperature $T_E$.
They are spaced at a distance $a\gg R$ from the graphene sheet which
is kept at the temperature $T_g$. In line with the Clausius-Mossotti
equation, the polarizabilities of dielectric and metallic nanoparticles
are given by
\begin{equation}
\alpha_0=R^3\frac{\ve-1}{\ve+2},\qquad
\alpha_0=R^3,
\label{eq1}
\end{equation}
\noindent
respectively, where in the separation region considered below one
can employ the static dielectric permittivity $\ve$  of a
dielectric material (the Gaussian system of units is used).
It is assumed also that $R\ll\lambda_{E,g}\equiv\hbar c/(k_BT_{E,g})$ \cite{20a}.
Note that for $T_E=300~$K we have $\lambda_E\approx 7.6~\mu$m.

The nonequilibrium Casimir-Polder force acting on a nanoparticle on the source
side of graphene is represented as a sum of the term which is often referred
to in the literature as ``equilibrium" and the proper
nonequilibrium addition to it \cite{8,10}
\begin{equation}
\nf\at=F\at+\Delta F\at.
\label{eq2}
\end{equation}
\noindent
Explicit expressions for both contributions on the r.h.s. of this equation
presented below were derived under the condition that our system is in local
thermal equilibrium, i.e., the temperatures of a nanoparticle and a graphene
sheet are constant, as well as the temperature of the environment.
The sphere radius and the separation distance between a sphere and a plate
satisfy the conditions formulated above.

The  term $F$ in Eq.~({\ref{eq2}) is  equal to half a sum of the  Casimir-Polder
forces
\begin{equation}
F\at=\frac{1}{2}\left[
\tef(a,T_E;T_g)+\tef(a,T_g;T_g)\right],
\label{eq2a}
\end{equation}
\noindent
where the first
temperature argument in $\tef$
indicates the temperature of the environment and the
second relates to the graphene sheet. If the dielectric response of the plate
does not depend on $T$, the quantities $\tef$ are calculated by the standard
Lifshitz formula \cite{2,3,4,5} with the environmental temperatures equal
to $T_E$ and $T_g$, respectively.
If, however, the dielectric response is  temperature-dependent, one
should again calculate $\tef(a,T_E;T_g)$ and $\tef(a,T_g;T_g)$ using the
standard Lifshitz formula with $T_E$ and $T_g$ as the temperatures of the
environment but in both cases substitute there the reflection coefficients
on a graphene sheet  calculated at the graphene temperature $T_g$ \cite{12}.
As an example,
\begin{eqnarray}
&&
\hspace*{-2mm}
\tef(a,T_E;T_g)=-\frac{2k_BT_E\alpha_0}{c^2}
\sum_{l=0}^{\infty}{\vphantom{\sum}}^{\prime}\int_0^{\infty}\!\!\!\kb d\kb
e^{-2aq_l}
\label{eq3} \\
&&\hspace*{-1mm}
\times\left[(2q_l^2c^2-\xi_l^2)R_{\rm TM}(i\xi_l,\kb;T_g)-
\xi_l^2 R_{\rm TE}(i\xi_l,\kb;T_g)\right].
\nonumber
\end{eqnarray}

In Eq.~({\ref{eq3}), $k_B$ is the Boltzmann constant, the prime on the sum in $l$
divides the term with $l=0$ by 2, $\kb$  is the magnitude of the wave vector
projection on the
plane of graphene, $\xi_l=2\pi k_BT_E l/\hbar$ with $l=0,\,1,\,2,\,\ldots$ are the
Matsubara frequencies calculated at the temperature $T_E$,
$q_l^2=\skb+\xi_l^2/c^2$, and $R_{\rm TM,TE}$ are the reflection coefficients on
a graphene sheet for the transverse magnetic, TM, and transverse electric, TE,
polarizations of the electromagnetic field, calculated at the temperature $T_g$.
It is assumed that an attractive force is negative.

Note that the force $\tef(a,T_g;T_g)$ is obtained from Eq.~(\ref{eq3}) by replacing
$T_E$ with $T_g$ in front of the sum and in the Matsubara frequencies, so that they
become equal to $\xi_l=2\pi k_BT_g l/\hbar$. The tilde in the used notations $\tef$
underlines that, although this quantity has the form of an equilibrium force,
its physical meaning is somewhat different. Thus, the proper equilibrium Casimir-Polder
force at the environmental temperature $T_E$ is defined as $\ef(a,T_E)=\tef(a,T_E;T_E)$
where $\tef$ is given in Eq.~(\ref{eq3}).
It is convenient to identically present the term $F\at$ in Eq.~(\ref{eq2a}) as
\begin{eqnarray}
&&
 F\at=\tef(a,T_E;T_g)
\label{eq4} \\
&&~~~~~~~
 +\frac{\tef(a,T_g;T_g)-\tef(a,T_E;T_g)}{2},
\nonumber
\end{eqnarray}
\noindent
where the first contribution on the r.h.s. is given in Eq.~(\ref{eq3}).
For our purposes, it is appropriate to write the second contribution
on the r.h.s. of Eq.~(\ref{eq4}) not
 according to Eq.~(\ref{eq3}) but using the equivalent representation
 of the Casimir-Polder force as an  integral along the  real frequency axis
 \cite{4,5}
\begin{eqnarray}
&&
\hspace*{-0.5cm}
\frac{\tef(a,T_g;T_g)-\tef(a,T_E;T_g)}{2}=\frac{\hbar\alpha_0}{\pi c^2}\!\int_0^{\infty}
\!\!\!\!\Theta(\omega,T_E,T_g)
\label{eq5}\\
&&
\hspace*{-0.2cm}
\times\left\{\int_0^{\omega/c}\!\!\!\kb d\kb \,{\rm Im}\left[
e^{2ia\sqrt{\frac{\omega^2}{c^2}-\skb}}\sum_{\kappa} A_{\kappa}
R_{\kappa}(\omega,\kb;T_g)\right]\right.
\nonumber \\
&&
\hspace*{-0.1cm}
+\left.
\int_{\omega/c}^{\infty}\!\!\!\kb d\kb \,e^{-2a\sqrt{\skb-\frac{\omega^2}{c^2}}}\,
{\rm Im}\left[\sum_{\kappa} A_{\kappa}R_{\kappa}(\omega,\kb;T_g)\right]\right\}.
\nonumber
\end{eqnarray}

In this equation, $\kappa={\rm TM,TE}$ is
the index indicating the palarization state, the quantities $A_{\kappa}$
are equal to
\begin{equation}
A_{\rm TM}=2\skb c^2-\omega^2, \qquad
A_{\rm TE}=\omega^2,
\label{eq5a}
\end{equation}
\noindent
and
\begin{equation}
\Theta(\omega,T_E,T_g)=\left(e^{\frac{\hbar\omega}{k_BT_E}}-1\right)^{-1}-
\left(e^{\frac{\hbar\omega}{k_BT_g}}-1\right)^{-1}.
\label{eq6}
\end{equation}
\noindent
Note that the explicit expressions for two reflection coefficients $R_{\kappa}$ in Eq.~(\ref{eq5}) are presented below in Eq.~(\ref{eq9}).

The general expression for a nonequilibrium addition in Eq.~(\ref{eq2})
was obtained in Ref.~\cite{10} for an atom-plate interaction
under a condition of local thermal equilibrium.
For real and constant polarizabilities (\ref{eq1})
describing spherical nanoparticles
whose radius satisfies the conditions formulated above, it is given by
\begin{eqnarray}
&&
\Delta F\at=-\frac{\hbar\alpha_0}{\pi c^2}\int_0^{\infty}\!\!d\omega
\Theta(\omega,T_E,T_g)
\label{eq7} \\
&&
\hspace*{-3mm}
\times\left\{\int_0^{\omega/c}\!\!\!\kb d\kb \,{\rm Im}\left[
e^{2ia\sqrt{\frac{\omega^2}{c^2}-\skb}}\sum_{\kappa} A_{\kappa}
R_{\kappa}(\omega,\kb;T_g)\right]\right.
\nonumber \\
&&
\hspace*{-2mm}
-\left.
\int_{\omega/c}^{\infty}\!\!\!\kb d\kb \,e^{-2a\sqrt{\skb-\frac{\omega^2}{c^2}}}\,
{\rm Im}\left[\sum_{\kappa} A_{\kappa}R_{\kappa}(\omega,\kb;T_g)\right]\right\}.
\nonumber
\end{eqnarray}

It is seen that Eqs.~(\ref{eq5}) and (\ref{eq7}) contain the contributions from
both propagating waves, for which $\kb< \omega/c$, and evanescent waves, for which
$\kb\geqslant \omega/c$. In doing so, the contributions of the propagating
waves enter Eqs.~(\ref{eq5}) and (\ref{eq7}) with the opposite sign. Substituting
Eqs.~(\ref{eq4}), (\ref{eq5}) and (\ref{eq7}) in Eq.~(\ref{eq2}), for the
nonequilibrium Casimir-Polder force between a nanoparticle and a graphene sheet
one finally obtains
\begin{eqnarray}
&&
\hspace*{-0.4cm}
\nf\at=\tef(a,T_E;T_g)+\frac{2\hbar\alpha_0}{\pi c^2}\int_0^{\infty}
\!\!\!\!d\omega\Theta(\omega,T_E,T_g)
\nonumber\\[-2mm]
&&
\label{eq8} \\[-2mm]
&&
\hspace*{-0.3cm}
\times
\int_{\omega/c}^{\infty}\!\!\!\kb d\kb \,e^{-2a\sqrt{\skb-\frac{\omega^2}{c^2}}}\,
{\rm Im}\left[\sum_{\kappa} A_{\kappa}R_{\kappa}(\omega,\kb;T_g)\right],
\nonumber
\end{eqnarray}
where $\tef(a,T_E;T_g)$ is given by Eq.~(\ref{eq3}).

It is pertinent to note that in the representation (\ref{eq8}) the nonequilibrium
contribution is determined by only the evanescent waves whereas the propagating
waves contribute to  $\nf$ implicitly only through $\tef(a,T_E;T_g)$.

\section{Electromagnetic response of graphene in terms of the polarization tensor}

The explicit expression for the polarization tensor of graphene $\Pi_{ij}\pot$,
$i,\,j=0,\,1,\,2$ valid over the entire plane of complex frequencies $\omega$
was found in Refs.~\cite{51,52} in the framework of the Dirac model applicable
at energies below 3~eV (the absorption peak of graphene at the wavelength of
270~nm corresponds to a higher energy of $\hbar\omega=4.59~$eV).
Note that the characteristic energy of the Casimir-Polder interaction
$\hbar\omega_c=\hbar c/(2a)$ at $a>100~$nm is below 1~eV. Therefore, at these
separations the description of the electromagnetic response of graphene using
the results of Refs.~\cite{51,52} is well founded.

It is convenient to express the reflection coefficients on a graphene sheet
via $\Pi_{00}$ and the combination of the tensor components
\begin{equation}
\Pi=\skb\Pi_i^{\,i}-q^2\Pi_{00},
\label{eq8a}
\end{equation}
\noindent
 where
\begin{equation}
 q^2\equiv q^2\ok=\skb-\frac{\omega^2}{c^2}.
\label{eq8b}
\end{equation}
\noindent
In terms of these quantities the reflection coefficients are given by
\cite{25,32,50,56}
\begin{eqnarray}
&&
R_{\rm TM}(\omega,\kb;T)=\frac{q\Pi_{00}\pot}{q\Pi_{00}\pot+2\hbar\skb},
\nonumber \\
&&
R_{\rm TE}(\omega,\kb;T)=-\frac{\Pi\pot}{\Pi\pot+2\hbar\skb q}.
\label{eq9}
\end{eqnarray}

General expressions for the components of the polarization tensor at real
$\omega$ are obtained in Ref.~\cite{51}. Here, we explicitly specify the form
of $\Pi_{00}$ and $\Pi$ in the region of evanescent waves $\kb>\omega/c$
used in Eq.~(\ref{eq8}). It is convenient to present $\Pi_{00}$ and $\Pi$ as
\begin{eqnarray}
&&
\Pi_{00}\pot=\Pi_{00}^{(0)}\ok+\Pi_{00}^{(1)}\pot,
\nonumber \\
&&
\Pi\pot=\Pi^{(0)}\ok+\Pi^{(1)}\pot,
\label{eq10}
\end{eqnarray}
\noindent
where the first terms on the right-hand sides indicate the zero-temperature
contribution and the second ones --- the thermal correction.
In the plasmonic region \cite{57}
$\omega/c<\kb\leqslant\omega/v_{\rm F}\approx 300\omega/c$ it holds \cite{51}
\begin{equation}
\Pi_{00}^{(0)}\ok=i\frac{\pi\alpha\hbar\skb}{p},
\quad
\Pi^{(0)}\ok=-i{\pi\alpha\hbar\skb}{p},
\label{eq11}
\end{equation}
\noindent
where $\alpha=e^2/(\hbar c)$ is the fine structure constant and
\begin{equation}
p^2\equiv p^2\ok=\frac{\omega^2-v_{\rm F}^2\skb}{c^2}.
\label{eq11a}
\end{equation}
\noindent
 These quantities are
pure imaginary.

The quantities $\Pi_{00}^{(1)}$  and $\Pi^{(1)}$ in the plasmonic region
are more complicated. They have nonzero real and imaginary parts. Thus,
from the results of Ref.~\cite{51} one obtains
\begin{equation}
{\rm Re}\Pi_{00}^{(1)}\pot=\frac{8\alpha\hbar c^2}{v_{\rm F}^2}
(I_1+I_2+I_3),
\label{eq12}
\end{equation}
where
\begin{eqnarray}
&&
I_1=2\int_0^{u^{(-)}}\!\!\frac{du}{e^{\beta u}+1}\left[1-\frac{1}{2cp}
\sum_{\lambda=\pm1}B_1(2cu+\lambda\omega)\right],
\nonumber \\
&&
I_2=2\int_{u^{(-)}}^{u^{(+)}}\!\!\frac{du}{e^{\beta u}+1}\left[1-\frac{1}{2cp}
B_1(2cu+\omega)\right],
\label{eq13} \\
&&
I_3=2\int_{u^{(+)}}^{\infty}\!\!\frac{du}{e^{\beta u}+1}\left[1-\frac{1}{2cp}
\sum_{\lambda=\pm1}\lambda B_1(2cu+\lambda\omega)\right].
\nonumber
\end{eqnarray}
\noindent
Here, $\beta=\hbar c/(k_BT)$, $u^{(\pm)}=(\omega\pm v_{\rm F}\kb)/(2c)$,
and $B_1(x)=(x^2-v_{\rm F}^2\skb)^{1/2}$.

The imaginary part of $\Pi_{00}^{(1)}$ is given by
\begin{equation}
{\rm Im}\Pi_{00}^{(1)}\pot=-\frac{8\alpha\hbar c}{v_{\rm F}^2p}
\int_{u^{(-)}}^{u^{(+)}}\!\!\!\!\!du
\frac{\sqrt{v_{\rm F}^2\skb-
(2cu-\omega)^2}}{e^{\beta u}+1}.
\label{eq14}
\end{equation}

In a similar way, from Ref.~\cite{51} after some identical transformations, for
the real part of $\Pi^{(1)}$ we find
\begin{equation}
{\rm Re}\Pi^{(1)}\pot=\frac{8\alpha\hbar \omega^2}{v_{\rm F}^2}
(J_1+J_2+J_3),
\label{eq15}
\vspace*{2mm}
\end{equation}
where
\begin{eqnarray}
\vspace*{2mm}
&&
J_1=2\int_0^{u^{(-)}}\!\!\frac{du}{e^{\beta u}+1}\left[1-\frac{cp}{2\omega^2}
\sum_{\lambda=\pm1}B_2(2cu+\lambda\omega)\right],
\nonumber \\
&&
J_2=2\int_{u^{(-)}}^{u^{(+)}}\!\!\frac{du}{e^{\beta u}+1}\left[1-\frac{cp}{2\omega^2}
B_2(2cu+\omega)\right],
\label{eq16} \\
&&
J_3=2\int_{u^{(+)}}^{\infty}\!\!\frac{du}{e^{\beta u}+1}\left[1-\frac{cp}{2\omega^2}
\sum_{\lambda=\pm1}\lambda B_2(2cu+\lambda\omega)\right].
\nonumber
\end{eqnarray}
\noindent
Here, $B_2(x)=x^2/(x^2-v_{\rm F}^2\skb)^{1/2}$.

The imaginary part of $\Pi^{(1)}$ takes the form
\begin{equation}
{\rm Im}\Pi^{(1)}\pot=\frac{8\alpha\hbar cp}{v_{\rm F}^2}
\nonumber
\end{equation}
\begin{equation}
\times
\int_{u^{(-)}}^{u^{(+)}}\!\!\!\!
\frac{du}{e^{\beta u}+1}\frac{(2cu-\omega)^2}{\sqrt{v_{\rm F}^2\skb-
(2cu-\omega)^2}}.
\label{eq17}
\end{equation}
\noindent
This concludes the consideration of the plasmonic region.

In the remaining region of evanescent waves $\kb>\omega/v_{\rm F}\approx 300\omega/c$
in Eq.~(\ref{eq8}), one finds \cite{51}
\begin{eqnarray}
&&
\Pi_{00}^{(0)}\ok=\frac{\pi\alpha\hbar\skb}{\tilde{p}}, \quad
\Pi^{(0)}\ok={\pi\alpha\hbar\skb}{\tilde{p}},
\label{eq18}
\\
&&
{\rm where}\qquad
\tilde{p}^2\equiv\tilde{p}^2\ok=
\frac{v_{\rm F}^2\skb-\omega^2}{c^2}.
\label{eq18a}
\end{eqnarray}
\noindent
These quantities are real.

The form of thermal corrections to the zero-temperature results (\ref{eq18})
 also  follows from the general expressions presented in Ref.~\cite{51},
 where $D=\hbar c\tilde{p}/(2k_BT)$. Note that the quantities (\ref{eq19})
are complex.
They have both real and imaginary parts.
 \begin{widetext}
\begin{eqnarray}
&&
\Pi_{00}^{(1)}\pot=\frac{8\alpha\hbar c^2\tilde{p}}{v_{\rm F}^2}\int_0^{\infty}
\!\!\frac{dv}{e^ {Dv}+1}\left[1+\frac{1}{2}\sum_{\lambda=\pm 1}\lambda
\left(1-v^2-\frac{2\lambda\omega}{c\tilde{p}}v\right)^{1/2}\right],
\nonumber \\
&&
\Pi^{(1)}\pot=\frac{8\alpha\hbar\omega^2\tilde{p}}{v_{\rm F}^2}\int_0^{\infty}
\!\!\frac{dv}{e^ {Dv}+1}\left[1+\frac{1}{2}\sum_{\lambda=\pm 1}\lambda
\frac{\left(\frac{c\tilde{p}v}{\omega}+\lambda\right)^2}{\left(1-v^2-
\frac{2\lambda\omega}{c\tilde{p}}v\right)^{1/2}}\right],
\label{eq19}
\end{eqnarray}
\end{widetext}
\noindent

Using Eqs.~(\ref{eq9})--(\ref{eq19}), one can compute the second, nonequilibrium,
contribution to the Casimir-Polder force in Eq.~(\ref{eq8}). As to the equilibrium
contribution, $\tef(a,T_E;T_g)$, it can be computed by Eq.~(\ref{eq3})
where the reflection
coefficients (\ref{eq9}) are calculated at the pure imaginary Matsubara frequencies
$\omega=i\xi_l$. In doing so the expressions for $\Pi_{00}$ and $\Pi$
at $\omega=i\xi_l$
are well known. They can be found in Ref.~\cite{58}. Numerically the same
values of $\Pi_{00}$ and $\Pi$ at $\omega=i\xi_l$ are obtained using the alternative
representation for the polarization tensor of Ref.~\cite{50} (this representation
was used in Refs.~\cite{25,26,27,29,60,61}).

\section{Computational results for the attractive and repulsive forces}

\begin{figure}[!b]
\vspace*{-1.5cm}
\centerline{\hspace*{-1.cm}
\includegraphics[width=4.5in]{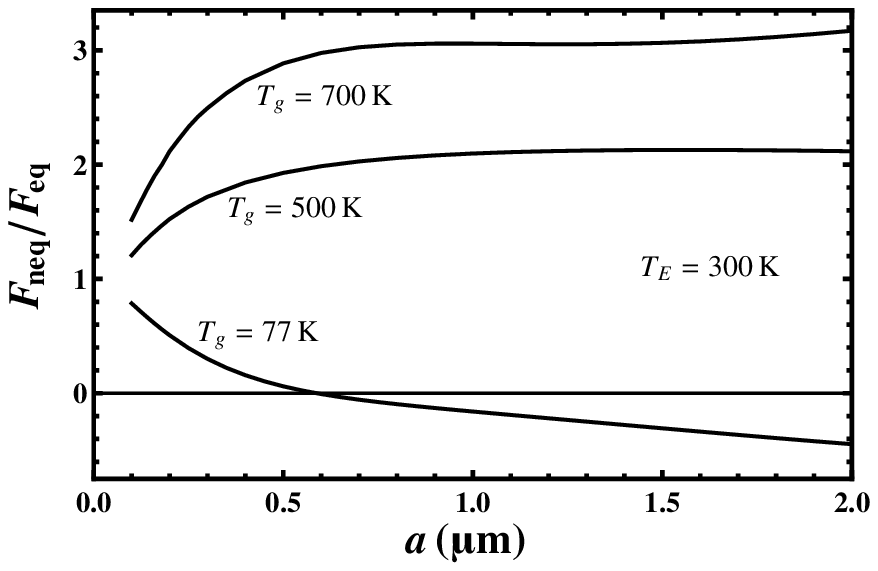}}
\vspace*{-9.cm}
\caption{\label{fg1}  The ratio of nonequilibrium to equilibrium
Casimir-Polder forces between a nanoparticle and a graphene sheet
is shown as the function of separation by the three lines for different
graphene temperatures $T_g$. In all cases the temperatures of a nanoparticle
and of the environment are equal to the temperature at thermal
equilibrium $T_E$.}
\end{figure}
First we compute the ratio of nonequilibrium Casimir-Polder force acting between
a nanoparticle and a graphene sheet, $\nf\at$, to the equilibrium one, $\ef(a,T_E)$.
These computations are performed by Eq.~(\ref{eq8}) and by Eq.~(\ref{eq3}) with
$T_E=T_g=300~$K. The computational results are presented in Fig.~\ref{fg1} as the
functions of nanoparticle-graphene separation by the top, middle and bottom lines
for the graphene temperature $T_g$ equal to 700~K, 500~K, and 77~K, respectively.
These results are valid for both dielectric and metallic nanoparticles of any
diameter $d=2R\ll a$ because the ratio under consideration does not depend on
$\alpha_0$.

As is seen in Fig.~\ref{fg1}, the effects of nonequilibrium increase the magnitude
of the total Casimir-Polder  force for $T_g>T_E$ and decrease it for $T_g<T_E$.
In the latter case, the total nonequilibrium force vanishes at some separation
and changes its sign at larger separations.
Thus, for a graphene sheet kept at 77~K the Casimir-Polder force vanishes at
$a\approx 0.58~\mu$m and becomes repulsive at $a> 0.58~\mu$m.
{}From Fig.~\ref{fg1} it is also seen that with increasing separation an
impact of the nonequilibrium effects becomes smaller.

\begin{figure}[!b]
\vspace*{-1.5cm}
\centerline{\hspace*{-1.cm}
\includegraphics[width=4.5in]{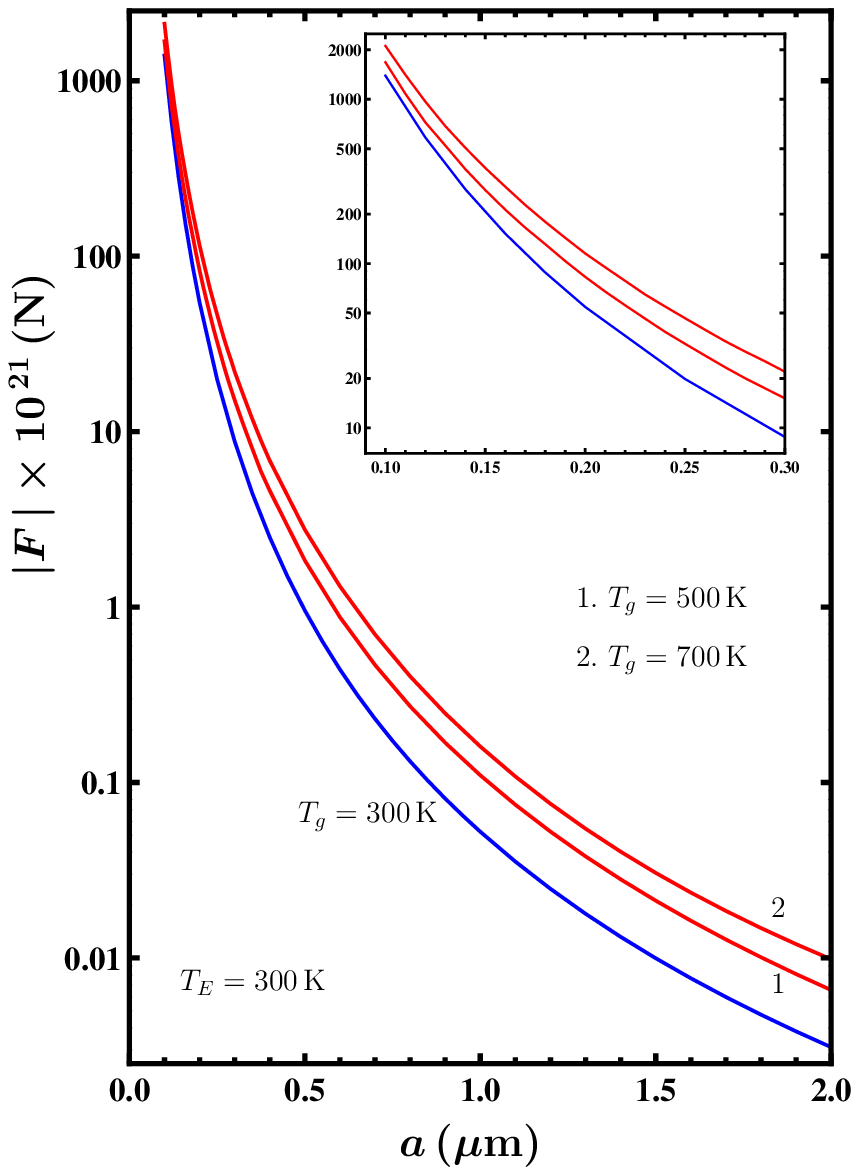}}
\vspace*{-3.5cm}
\caption{\label{fg2}  The magnitude of nonequilibrium Casimir-Polder force between a metallic
nanoparticle of diameter 5\,nm and a heated graphene sheet is shown
as the function of separation by the lines 1 and 2 for two different
graphene temperatures $T_g$. The respective equilibrium force is shown
by the bottom line. The region of short separations is shown in the inset
on an enlarged scale. }
\end{figure}
In Fig.~\ref{fg2}, the computational results for the magnitude of the Casimir-Polder
force, $\nf\at$, between a metallic nanoparticle of $d=5~$nm diameter and a graphene
sheet are presented as the functions of separation in a logarithmic scale by the
lines labeled 1 and 2 for $T_g=500~$K and 700~K, respectively, in comparison
with the equilibrium force, $\ef(a,T_E)$ calculated with
$T_g=T_E=300~$K, shown by the bottom line.
In the inset, the region of short nanoparticle-graphene separations is shown
on an enlarged scale. All force in Fig.~\ref{fg2} are negative which corresponds
to attraction. It is seen  that with increasing temperature the magnitude of the
Casimir-Polder force increases which leads to stronger attraction.

Similar results for the case of graphene sheet cooled down to $T_g=77~$K are shown
in Fig.~\ref{fg3}. Here, the bottom line demonstrates the magnitude of the total
nonequilibrium Casimir-Polder force, $\nf\at$, which vanishes at $a\approx 0.8~\mu$m
and becomes repulsive at larger separations. The top line reproduces the bottom line of
Fig.~\ref{fg2} which shows the equilibrium Casimir-Polder force, $\ef(a,T_E)$,
computed with $T_g=T_E=300~$K. The inset again presents the region of short separations
on an enlarged scale. Although the force magnitude at  $T_g=T_E=300~$K is again larger
than at $T_g=77~$K, the qualitatively new effect arising at $T_g<T_E$ is the change of
attraction with repulsion for sufficiently large separations between a nanoparticle
and a graphene sheet.
\begin{figure}[!h]
\vspace*{-4.cm}
\centerline{\hspace*{-1.cm}
\includegraphics[width=4.5in]{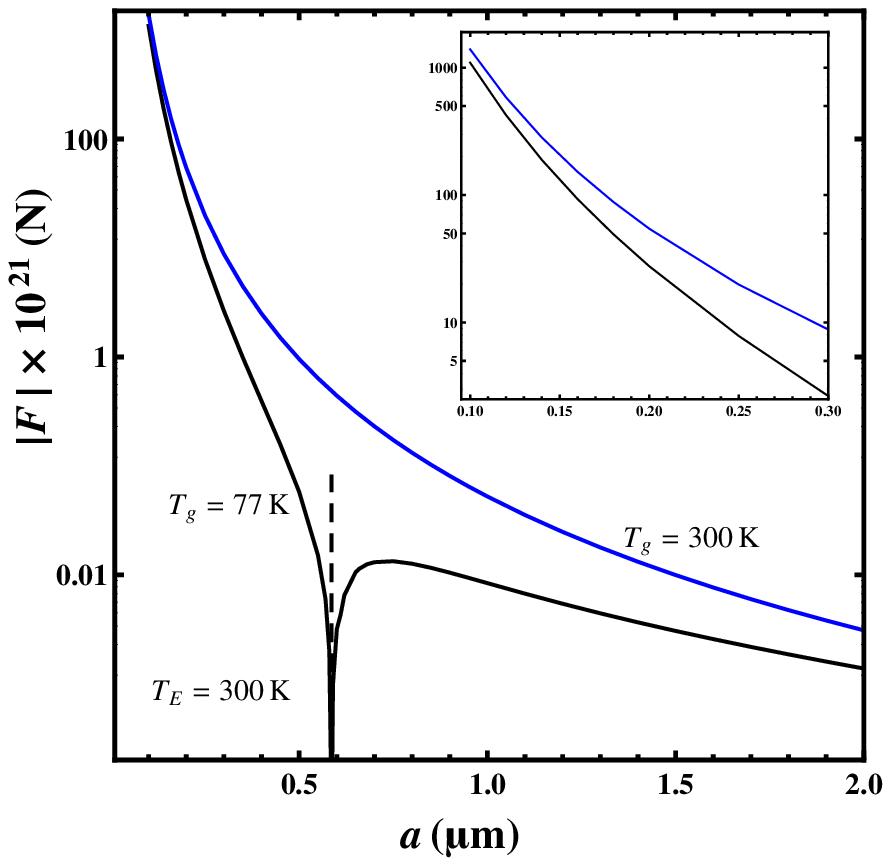}}
\vspace*{-3.5cm}
\caption{\label{fg3}   The magnitude of nonequilibrium Casimir-Polder force between a metallic
nanoparticle of diameter 5\,nm and a cooled graphene sheet is shown as
the function of separation by the bottom line. The force is attractive to the
left of the dashed line and repulsive to the right of it. The respective
equilibrium force is shown by the top line. The region of short separations
is shown in the inset on an enlarged scale.}
\end{figure}

\section{Conclusions and outlook}

In the foregoing, we have considered
the Casimir-Polder interaction between a spherical nanoparticle and
a graphene sheet in out-of-thermal-equilibrium conditions when the
common temperature of a nanoparticle and of the environment can be
different from the temperature of graphene. This problem is of
interest for both fundamental and applied physics because the
electromagnetic response of graphene strongly depends on temperature.
The obtained expression for the Casimir-Polder force between a
nanoparticle and a graphene sheet is based on the theory of atom-wall
interaction out of thermal equilibrium \cite{6,7,8,9,10,11} adapted
for the case of temperature-dependent dielectric response in
Ref. \cite{12}. In so doing, the response function of graphene is
described by means of the polarization tensor in the framework of the
Dirac model.

In accordance to the results obtained, the magnitude of the
Casimir-Polder force on a nanoparticle increases with increasing
temperature of a graphene sheet although an impact of the nonequilibrium
conditions decreases with increasing separation. An important qualitative
effect is that at some separation distance the attractive
nanoparticle-graphene force vanishes if the graphene temperature is
lower than the temperature of the environment and becomes repulsive at
larger separations.
For a small sphere interacting with a dielectric plate characterized
by the temperature-independent permittivity function, similar nonequilibrium
effects were obtained in Ref.~\cite{20a} at separation distances of a few
micrometers. According to our results, in the case of graphene the effects
of thermal nonequilibrium become essential at much shorter separation
distances. This opens novel opportunities for the
experimental investigation of graphene systems and for the control of
forces between nanoparticles and graphene sheet with the aim of
applications in bioelectronics.

\section*{Acknowledgments}
\vspace*{-0.4cm}
The work of O.~Yu.~T. was supported by the
Russian Science Foundation under Grant No.\ 21-72-20029. G.~L.~K.\ and
V.~M.~M.\ were partially supported by the Peter the Great Saint
Petersburg Polytechnic University in the framework of the Russian state
assignment for basic research (Project No.\ FSEG-2020-0024). The work
of V.~M.~M.\ was also supported by the Kazan Federal University
Strategic Academic Leadership Program.


\end{document}